\documentclass[apl,twocolumn,superscriptaddress]{revtex4-1}
\usepackage[latin1]{inputenc}
\usepackage{amsmath}
\usepackage{amsfonts}
\usepackage{amssymb}
\usepackage{graphicx}
\usepackage{float}
\begin{document}

\title{Determination of surface and interface magnetic properties
for the multiferroic heterostructure Co/BaTiO$_3$ using SPLEED and ARPES}
\date{\today}
\author{Stephan Borek}
\author{J\"urgen Braun}
\affiliation{Department Chemie, Ludwig-Maximilians-Universit\"{a}t M\"{u}nchen, Butenandtstra\ss e 5-13, 81377 M\"{u}nchen, Germany}
\author{J\'an Min\'ar}
\affiliation{Department Chemie, Ludwig-Maximilians-Universit\"{a}t M\"{u}nchen, Butenandtstra\ss e 5-13, 81377 M\"{u}nchen, Germany}
\affiliation{New Technologies-Research Centre, University of West Bohemia,\\ Univerzitni 8, 306 14 Pilsen, Czech Republic}
\author{Dimitry Kutnyakhov}
\affiliation{Johannes-Gutenberg-Universit\"{a}t Mainz, Staudinger Weg 7, 55128 Mainz, Germany}
\author{Hans-Joachim Elmers}
\affiliation{Johannes-Gutenberg-Universit\"{a}t Mainz, Staudinger Weg 7, 55128 Mainz, Germany}
\author{Gerd Sch\"onhense}
\affiliation{Johannes-Gutenberg-Universit\"{a}t Mainz, Staudinger Weg 7, 55128 Mainz, Germany}
\author{Hubert Ebert}
\affiliation{Department Chemie, Ludwig-Maximilians-Universit\"{a}t M\"{u}nchen, Butenandtstra\ss e 5-13, 81377 M\"{u}nchen, Germany}

\begin{abstract}
Co/BaTiO$_3$(001) is one of the most interesting multiferroic heterostructures as it combines different ferroic phases,
setting this way the fundamentals for innovative technical applications. Various theoretical approaches have been applied 
to investigate the electronic and magnetic properties of Co/BaTiO$_3$(001). Here we determine the magnetic properties of
3 ML Co/BaTiO$_3$ by calculating spin-polarized electron diffraction as well as angle-resolved photoemission spectra,
with both methods being well established as surface sensitive techniques. Furthermore, we discuss the impact of altering
the BaTiO$_3$ polarization on the spectra and ascribe the observed changes to characteristic details of the
electronic structure.
\end{abstract}

\maketitle

\section{Introduction}

The combination of two ferroic phases may lead to new materials with interesting behaviour and corresponding applications
which are based on the electronic and magnetic properties of their constituents. This special class of materials is named
multiferroics and was introduced first in the 1960's by Smolenskii and Venevtsev \cite{SC82,VG94}.
In 1994 the interest in this topic has been renewed by H. Schmid \cite{schmidt1}, who introduced first the synomym
multiferroics.

For this class of materials one has to distinguish between the combination of two or more ferroic properties in one single
phase and the so-called multiferroic heterostructures. The latter ones are very interesting for technical applications
because their material properties can be tuned by shaping the interface between the ferroic phases according to the
technical requirements. An essential step in realizing such systems was the tremendous progress in developing crystal grow
techniques which allow the creation of defined interfaces \cite{dawber1}. Among the various multiferroic heterostructures
which have been studied so far one promising candidate is the Co/BaTiO$_3$ (Co/BTO) system \cite{HBM+15}. In contrast to the
well-established multiferroic heterostructure 3 ML Fe/BTO the 3 ML Co/BTO system has the advantage of ferromagnetic stability at room temperature
\cite{HBM+15}.

In a recent publication it was shown that the interface of Co/BTO is very similar to that of the intensively studied Fe/BTO \cite{HBM+15}.
It was discovered in detail that the coupling of the involved ferroic phases is caused by nearest neighbor interaction at the
interface. For the investigated systems of 1, 2 and 3 ML Co/BTO the strongest coupling was predicted for 2 ML Co. Similar to the
multiferroic heterostructure Fe/BTO the system Co/BTO opens the possibility to control the magnetic properties of the Co
layer changing the electric polarization of BTO. The ferroelectric polarization could be altered using external voltage or stress.
On the other hand it is possible to change the magnetic moments (direction or magnitude) of the Co layers affecting this way the
ferroelectric polarization \cite{fechner1}.

For the various properties predicted by theory a corresponding verification by experiment
is mandatory.
A well established way to do so is the
use of spectroscopy. For example, it was shown that using x-ray magnetic circular dichroism (XMCD) and x-ray
magnetic linear dichroism (XMLD), it is possible to detect the altering of the BTO polarization by investigating the magnetic
properties of Co \cite{HBM+15,BMF+12}. By comparing both methods it was shown that XMLD reacts more sensitive on a change of
the electrical polarization. However, there exist other suitable methods for studying magnetic and electronic properties at
surfaces, as for example, angle-resolved photoemission spectroscopy (ARPES) or spin-polarized low energy electron diffraction
(SPLEED). In this work we performed first principles calculation for
both methods to investigate the surface magnetic properties
of 3 ML Co/BTO affected by an altering of the BTO polarization. The surface sensitive SPLEED technique allows for a somewhat indirect
analysis of the coupling mechanisms at the interface. ARPES on the other hand enables a detailed investigation of the surface and interface
electronic structure. Especially changes of the dispersion relation which occur during a polarization change of the BTO can
be quantitatively monitored. To account for the high surface sensitivity of both methods we used a fully relaxed surface
and interface structure for 3 ML Co/BTO. Our calculations have been done by use of  fully relativistic multiple scattering
techniques in the framework of spin-polarized density functional theory \cite{SPR-KKR6.3}. Therefore, all effects originating
from spin-orbit coupling and exchange interaction are treated on the same level of accuracy \cite{ebert1,SPR-KKR6.3}.

The paper is organized as follows: In Sec. \ref{sec_theory} we briefly describe the theoretical methods used in this work,
i.e. the SPLEED and ARPES formalism. In Sec. \ref{sec_discussion} we discuss our theoretical results and in Sec. \ref{sec_summary}
we summarize our findings.

\section{Theoretical application \label{sec_theory}}

As described in our previous work we use the layer-KKR approach to the calculate the corresponding SPLEED pattern
\cite{feder2,BBM+15}.  Within the scope of this method a semi-infinite surface system is treated as a stack
of atomic layers using the so-called layer doubling method \cite{Pendry_SPLEED}. For every specific atomic layer the multiple
scattering of the incident electron has to be determined. The potentials needed for this computational step have been taken
from previous works \cite{Bor13,HBM+15}. Based on these potentials the single-site scattering matrices can be evaluated,
which are needed for the calculation of the so called Kambe X-matrix \cite{Kambe}, which determines the multiple
scattering in a specific layer. In a last step inter-layer scattering has to be considered. With the determination of
both scattering mechanisms one can calculate the so-called bulk reflection matrix, which represents the scattering of an
electron from a semi-infinite stack of atomic layers \cite{feder1,feder2}.

The self-consistent calculation of the electronic structure has been done for a half-space which consists out of four unit cells
of BTO, i.e. beyond this bulk properties have been assumed.
With respect to a transition from the inner BTO potentials to the surface the electronic structure was
allowed to relax. This guarantees a realistic description of the electronic structure at the interface and surface.
Similar to our investigations done for the multiferroic heterostructure Fe/BTO the calculations are based on the
tetragonal distorted structure of BTO. In this phase (P4mmm) the BTO has a permanent electrical polarization
originating from a shift of the Ti and O atoms \cite{CBS+11}. For BTO we assumed a lattice constant of 3.943 \AA\,
\cite{Bor13}.

The relaxed crystal structure at the interface was determined using the VASP code \cite{KF96,KJ99,HBM+15}.
We applied the resulting crystal structure and the the corresponding self-consistent potentials as input quantities
for our fully relativistic multiple scattering formalism to calculate the density of states (DOS), the SPLEED pattern
and ARPES intensities \cite{SPR-KKR6.3}.

The scattering process itself includes different parameters. The most important one is the location of the scattering plane
with respect to the sample surface. The scattering plane is defined by the wave vectors of the incident and scattered electrons.
Additionally, the direction of the surface magnetization and the polarization of the incident electron have to be defined.
In our calculation both have been aligned perpendicular to the scattering plane. As can be shown by symmetry considerations
for this setup exchange and spin orbit scattering contribute both \cite{tamura1}. With respect to the used scattering
configuration one has to calculate four reflectivities for all combinations of the electron polarization and the surface
magnetization ($I_{\mu}^{\sigma}$) \cite{feder1,tamura1}. The indices $\mu$ and $\sigma$ represent the surface magnetization
direction and the electron polarization, respectively. Based on these quantities the effective reflectivity $I_{eff}$ can be calculated via:

\begin{equation}
I_{eff}=\frac{1}{4}(I_+^++I_+^-+I_-^++I_-^-)
\label{eq:effective_reflectivity}
\end{equation}

The reflectivities ($I_{\mu}^{\sigma}$) enable the calculation of additional observables measured in the experiment, i.e.
spin-orbit asymmetry, exchange asymmetry and figure of merit. The exchange asymmetry is defined as:

\begin{equation}
 A_{ex,+}=\frac{I_{+}^{+}-I_{+}^{-}}{I_{+}^{+}+I_{+}^{-}}. \label{eq:exchange_asymmetry}
\end{equation}

The plus index at $A_{ex,+}$ indicates here a fixed magnetization direction along [100]. For all following calculations
the magnetization was oriented along this direction.

Additional quantities which are important for the calculation of SPLEED patterns determine the escape of the electron
into the vacuum, i.e. the work function and the surface potential barrier. For the work function we assumed the value 4.7 eV, which
has shown to be reasonable for diffraction calculations on similar systems \cite{BBE+15}. This choice is close 
to the experimental value of 5.0 eV \cite{haynes}. Concerning the surface barrier we have taken a parameterization
based on Rundgren and Malmstr\"om \cite{skriver1,rundgren1}. This type of surface barrier was applied in the past successfully to a variety of transition
metal systems. Using the method described so far we have calculated SPLEED patterns for a wide range of kinetic energies
and polar angles for both polarization directions of BTO (P$_{\text{up}}$, P$_{\text{down}}$).

The ARPES calculations have been performed in the framework of the so-called one-step model \cite{Pendry_SPLEED,Braun}.
In contrast to the three-step model which treats the  electron emission in different steps in the electron emission,
i.e. excitation of the electron, travel of the electron through the solid to the surface, emission of the electron into
the vacuum, all excitation steps are explicitly included in the one-step model. Beside the three-step model neglects
several effects important for the theoretical description of the photoemission process. As for example the interference
of surface and bulk emission, the interaction of the photoelectron with atoms during the transport to the surface and
a proper description of the transition process of the electron into the vacuum. For a more detailed description
of the ARPES calculations see for example \cite{braun1,hueffner1}. We investigated a fully relaxed surface and interface structure with an 
(001) orientation of BTO. The crystal and electronic structure has been investigated in detail in previous works
\cite{HBM+15}. Here we summarize only some essential structural parameters for the 3 ML Co/BTO interface (see Tab.
\ref{crystal_structure}). In Fig. \ref{crystal_structure} the corresponding structure at the interface of the Co/BTO system
is shown schematically.

\begin{figure}[h]
\centering
\includegraphics[scale=0.25,angle=0,clip=false,trim=0cm 0cm 0cm 0cm]{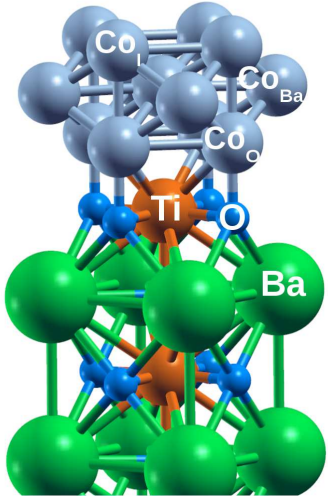}
\caption{Crystal structure of 3 ML Co/BTO. The non-equivalent
atomic sides have been indicated by name (not shown Co$\rm_{Ti}$
located on top of the Ti atom, color online).}
\label{fig:crystal_structure}
\end{figure}

\begin{table}[H]
\centering
\begin{tabular}{l|c c}
& P$\rm_{up}$ & P$\rm_{down}$ \\ 
\toprule
O-Co$\rm_O$ & 1.829 & 1.827 \\ 
I-II & 1.179 & 1.183 \\ 
II-III & 1.153 & 1.154 \\ 
Ti-Co$\rm_{Ti}$ & 3.014 & 3.094 \\ 
\end{tabular}
\caption{Atomic layer distances for the interface of 3 ML Co/BTO.
The distances are given in units of (\AA) \cite{HBM+15}.
Both polarization directions of BTO are shown. BTO polarization
pointing in surface direction (P$\rm_{up}$). BTO polarization
pointing in substrate direction (P$\rm_{down}$).
(I: Co layer on top of BTO, II: second Co layer,
III: third Co layer representing the surface).}
\label{crystal_structure}
\end{table}

The crystal structure of the Co/BTO system shows many similarities when compared to the Fe/BTO system, which was
investigated in previous studies \cite{fechner2,BMF+12,BBM+15}. The first Co layer (Co$_{\text{O}}$) is placed on
top of the O atoms of the Ti-O terminated BTO surface. In the second Co layer two non-equivalent Co positions occur.
The first position is on top of Ti (Co$_{\text{Ti}}$) and the second position is on top of Ba (Co$_{\text{Ba}}$).
According to previous studies the largest changes when switching the BTO polarization result for Ti at the interface 
inducing a large displacement of the Co$_{\text{Ti}}$ atom \cite{MFO}. Therefore, the interaction of the
Co$_{\text{Ti}}$ atom with the BTO substrate dominates the coupling between the ferroic materials and gives the
main contribution to changes in the electronic structure when altering the BTO polarization. The third Co
layer (Co$_{\text{I}}$) is placed on top of the interfacial O atoms. Concerning theses structural details 3 ML Co
form an unusual tetragonal distorted based centered crystal structure with electronic properties that are closely
related to 3 ML Fe/BTO \cite{fechner2}.

\section{Results and Discussion \label{sec_discussion}}

Due to the exchange scattering the magnetic properties at the Co surface are essential for our SPLEED calculations.
The magnetic moments at the surface are affected by the reduced number of nearest neighbors and by hybridization
effects with atoms at the interface. It has been shown that the magnetic properties depend on the number of Co
layers on top of BTO \cite{HBM+15}. For the systems investigated so far (1, 2, 3 ML Co/BTO) a ferromagnetic ground
state occurs. In actual investigations an in-plane configuration for 2 ML Co/BTO was predicted \cite{HBM+15}. Beside this it was
shown that the Co spin magnetic moments are strongly affected by the geometry of the Co films \cite{HBM+15}.
Especially, the Co spin magnetic moments are influenced by the hybridization of the Co 3$d$ states with states
of the substrates atoms. It was shown that the spin magnetic moment for 3 ML Co/BTO are quenched for all three
atomic layers which is based on the lower volume of the Co atoms according to the relaxed crystal structure
 \cite{HBM+15}.

\begin{figure}[ht]
\includegraphics[scale=0.35,angle=0,clip=true,trim=0cm 5cm 0cm 0cm]{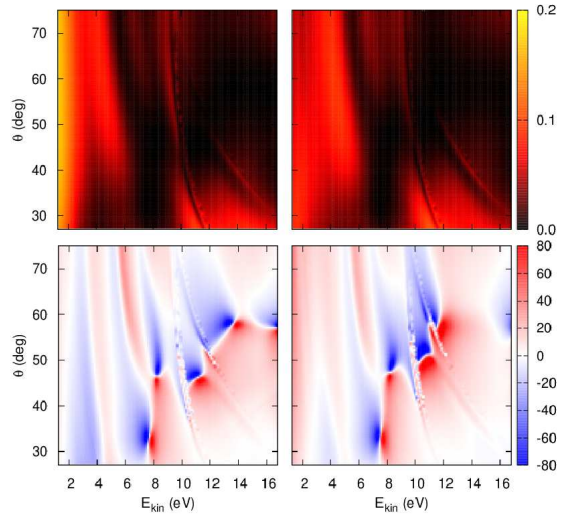}
\caption{SPLEED pattern for 3 ML Co/BTO. Reflectivity (top panel) and
exchange asymmetry (bottom panel, values given in (\%)).
The plane of incidence was aligned along the [010] direction
the magnetization of Co along [100].
Left: The polarization of the BTO is directed towards the surface (P$\rm_{up}$). 
Right: The polarization of the BTO pointing in opposite direction (P$\rm_{down}$) (color online).}
\label{spleed}
\end{figure}

In Fig. \ref{spleed} the SPLEED patterns for reflectivity (top row) and exchange asymmetry (bottom row) are shown.
Both quantities have been calculated for an upward (in surface direction) oriented polarization of BTO (left column)
and a downward oriented polarization of BTO (right column).

As shown in Fig. \ref{spleed} a dependency of the SPLEED pattern on the surface magnetization
(i.e. the BTO polarization) occurs. For kinetic energies less than 6 eV the reflectivity contains large
changes when switching the BTO polarization. This is in contrast to higher kinetic energies where the
reflectivity is less sensitive to a BTO altering. This result is related to the coupling of incident
electrons to bands which disperses perpendicular to the Co(001) surface. Depending on the band dispersion
along the normal vector of the (001)-plane and the kinetic energy of the incident electron a coupling is
possible or not \cite{BOC99}. Without coupling a large reflectivity results, with vanishing coupling
the intensity of the reflected electrons decreases.

The reflectivity pattern decompose in two main parts. Below 6 eV a high reflectivity occurs whereas
for higher kinetic energies the reflectivity decreases. In addition, below 6 eV switching the BTO polarization
has a higher impact on the effective reflectivity. This is caused by a higher sensitivity of electrons with
lower kinetic energy to changes in the magnetic properties at the surface. According to Fig. \ref{spleed},
electrons with kinetic energies above 8 eV can couple with states inside the crystal resulting in a decrease of the
effective reflectivity. In addition electrons with higher kinetic energy are less sensitivity to changes
of the surface magnetization. For the Co/BTO system one can therefore expect that both effects contribute to the
resulting SPLEED patterns.

In Fig. \ref{spleed} (bottom panel) the exchange asymmetries are presented. The exchange asymmetry gives a more
detailed picture with additional information concerning the magnetic properties. This is due to the loss of
information when calculating the effective reflectivity averaging over the individual spin-dependent electron beams. 
In the exchange asymmetry pattern, especially for higher kinetic energies (12-16 eV), pronounced changes in the polarization
are visible. Additionally, the exchange asymmetry alters for low kinetic energies ($\le$ 4 eV). Therefore, the
incident electron is affected by a change of the surface magnetization over the total kinetic energy range.

In principle the occurrence of an exchange asymmetry indicates an energetically split unoccupied band structure
\cite{OYY+08}. These exchange-split unoccupied bands are distinguishable via SPLEED. In Fig. \ref{spleed} the red
areas represent a parallel alignment of electron spin and surface magnetization whereas the blue areas stand for an
antiparallel alignment. For special regions with positive (red) or negative (blue) exchange asymmetry the incident
electrons can couple to majority or minority bands. According to the definition of the exchange asymmetry positive
values correspond to a reflection of the incident electron from the majority (spin up) states. For example, with
respect to kinetic energy and polar angle a parallel alignment of the electron spin and the surface magnetization
(red areas) results in a coupling of the incident electron to the minority bands whereas the coupling to the majority
bands is less pronounced. Therefore, the reflectivity of electrons with parallel aligned spin and surface magnetization
is higher than for an antiparallel alignment resulting in a positive exchange asymmetry. At band gaps and band
crossings the exchange asymmetry is zero. Based on this information it becomes extractable which spin projected
bands have been affected significantly by a change of the BTO polarization. In Fig. \ref{spleed} it can be seen that
for kinetic energies of 14 eV and a polar angle of 60 deg a change of the BTO polarization affects the exchange
asymmetry strongly. In this energy range the scattering is mainly due to majority bands (red color). For kinetic
energies of 12 eV the same is visible for scattering at minority bands (blue color). Due to the fact that the
splitting of the unoccupied states are related to the exchange splitting of the occupied states the SPLEED exchange
pattern gives a fingerprint of the magnetic properties at the surface influenced by the BTO polarization.

\begin{figure}[h]
\begin{minipage}[c]{.2\textwidth}
\includegraphics[scale=0.35,angle=270,clip=true,trim=0cm 5cm 0cm 0cm]{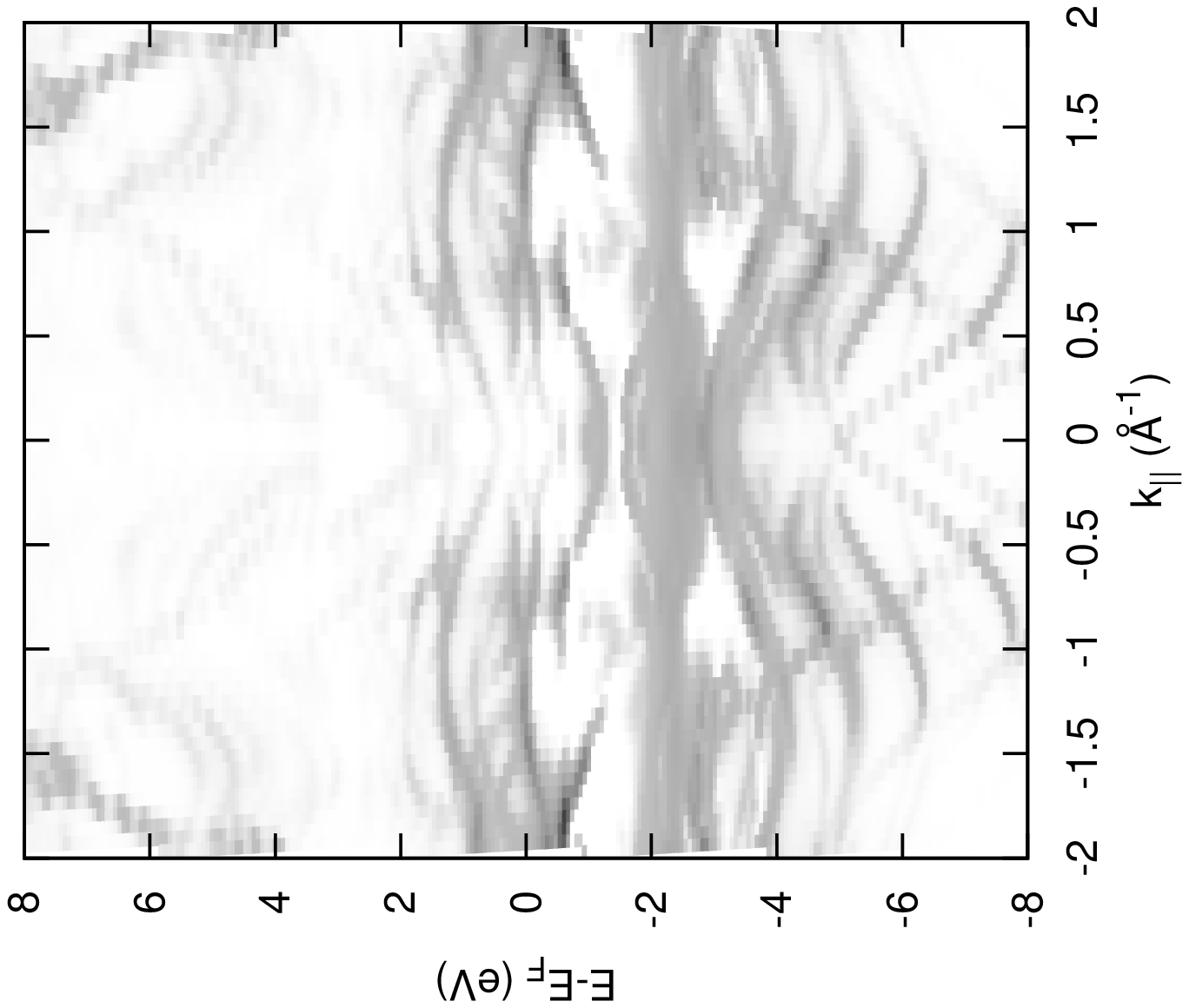}
\end{minipage}
\hspace{0.8cm}
\begin{minipage}[c]{.2\textwidth}
\includegraphics[scale=0.35,angle=270,clip=true,trim=0cm 7.2cm 0cm 0cm]{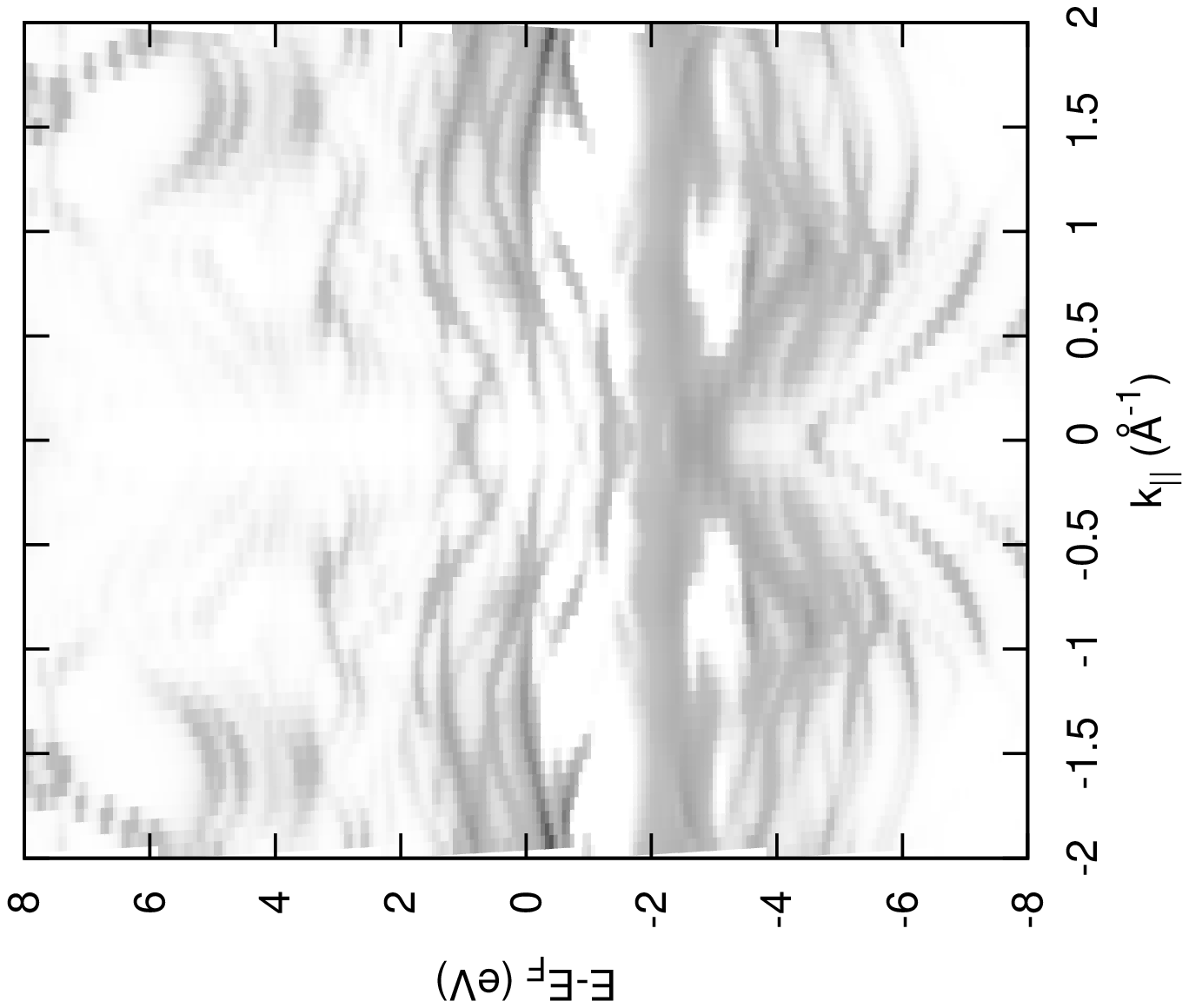}
\end{minipage}
\caption{One-step calculation of photoemission spectra for 3 ML Co/BTO.
The calculations assume normal incidence, a photon energy of 60 eV
and a [100] magnetization direction for the Co layers.
Left: The polarization of the BTO is directed along the surface normal (P$\rm_{up}$). 
Right: The polarization of the BTO is directed in opposite direction (P$\rm_{down}$).}
\label{pes}
\end{figure}

In addition we calculated angle-resolved photoemission spectra for 3 ML Co/BTO to visualize the impact
of the BTO polarization for a sample path along the $\overline{\text{H}}$-$\overline{\Gamma}$-$\overline{\text{H}}$
direction along the surface Brillouin zone. The corresponding results for the two polarization directions of the BTO
are shown in Fig. \ref{pes}.

The different polarizations at the interface strongly influence the band dispersion. The main changes in the
dispersion relation by altering the BTO polarization are visible at the Fermi energy. This reflects the changes
of the magnetic properties due to a coupling between Co and BTO. Below -5 eV the band dispersion includes bands
with quadratic $\textbf{k}$ dependence as expected from free electron-like behaviour. Above -5 eV the bands
show the typical dispersion of $d$ bands for transition metals, i.e. a small band width resulting in high
photoemission intensities.

\begin{figure}[h]
\center
\includegraphics[scale=0.40,angle=270]{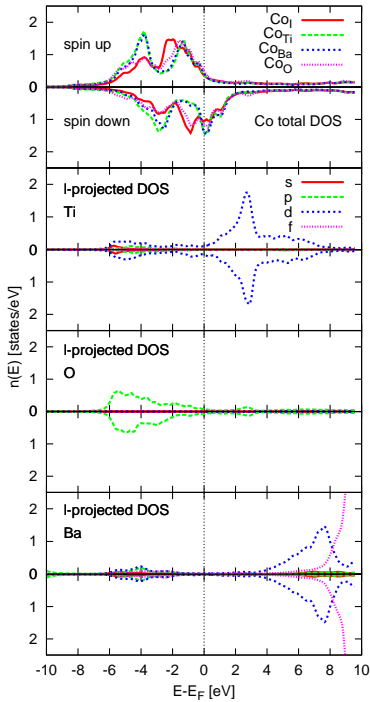}
\caption{DOS for Co/BTO(001) for the various atomic types near the interface.
The polarization of the BTO is directed along the surface normal (P$\rm_{up}$). (color online)}
\label{dos}
\end{figure}

The band dispersion is closely related to the effective mass of the electrons representing the interaction
with ion cores. For a large interaction the effective mass becomes large with respect to the free electron
mass resulting in a small bandwidth \cite{christman}. These band dispersions are visible above -5 eV. 
Additionally, to the ARPES band dispersions we calculated the DOS for the Co/BTO system. In Fig. \ref{dos} we
plotted the results for atoms at the interface (BTO P$\rm_{up}$).

Comparing the calculated ARPES spectra (left side, P$\rm_{up}$) with the plotted DOS it is visible that the main
contribution to the bands in the energy range from -2 eV to -4 eV originates from O $p$ states, Co $d$ states and a small
contribution of Ti $d$ states. A hybridization between these states is visible. In contrast the bands at the
Fermi level are mainly Co $d$ states. The appearance of Co $d$ states from -6 eV to 1 eV is due to the exchange
splitting, i.e. the spin magnetic moment of Co. With respect to Fig.~\ref{dos} the main contribution to the
unoccupied states arises from $d$ states of Ti ($\approx$ 3 eV) and $d$, $f$ states of Ba ($\ge$ 6 eV). It is also
visible that the magnetic properties of the Co/BTO system can be related to the Co atoms where for Ti and O due
to hybridization effects a small magnetic moment is induced \cite{HBM+15}. As has been shown in our previous work on
Fe/BTO it is reasonable to assume that the incident electron interacts only with the first atomic layer of Co.
Therefore, the induced magnetic moments for Ti and O can be neclegted concerning the interpretation of the
characteristic structures in the SPLEED pattern.

In previous works the coupling mechanism at the interface was investigated
via XMCD and XMLD \cite{HBM+15,Bor13,BMF+12}.
Similar to these methods SPLEED probes the unoccupied states \cite{OYY+08}.
In contrast to XAS where the transitions involve electronic states just above the Fermi energy in SPLEED the electrons have
to eject into the vacuum passing the surface potential. In Fig. \ref{pes} this corresponds to an energy of
4.7 eV above the Fermi level. As visible in the ARPES spectra above 5 eV large changes occur in the band
dispersion when switching the BTO polarization. This is the origin of the changes in the exchange asymmetry
enabling a high sensitivity for the investigation of the ferroic coupling at the interface.

\section{Summary \label{sec_summary}}

We have used a fully relativistic multiple-scattering technique to calculate SPLEED pattern and ARPES intensities
for the multiferroic system 3 ML Co/BTO(001). We have investigated the impact of switching the BTO polarization on
the exchange scattering and on the resulting ARPES spectra. We have shown that both methods are sensitive to 
the BTO polarization, giving the opportunity for experimental investigations on the coupling mechanisms for this multiferroic
heterostructure. 

\section{Acknowledgement}
We thank the BMBF (05K13WMA), the DFG (FOR 1346),
CENTEM PLUS (LO1402),
and the COST Action MP 1306 for financial support.
SB thank the SFB 762: Functionality of Oxide Interfaces
(Prof. I. Mertig, Dr. habil. A. Chass\'e)
providing the framework for the calculation of
the self-consistent potentials.

\bibliographystyle{aipnum4-1}
\bibliography{spleed_co_bto_new}

\end{document}